\documentclass[prl,aps,twocolumn,superscriptaddress]{revtex4-2}
\usepackage{graphicx,color}
\usepackage{amsthm}
\usepackage{amsfonts}
\usepackage{algorithmic}
\usepackage{enumerate}
\usepackage{latexsym}
\usepackage{amsmath}
\usepackage{amssymb}
\usepackage{subfig}
\usepackage[colorlinks=true,citecolor=blue,linkcolor=blue]{hyperref}

\usepackage{dcolumn}
\usepackage{bm}

\usepackage[T1]{fontenc}
\usepackage{mathptmx}

\begin{document}

\title{Charge stripe and superconductivity tuned by interlayer interaction in  a sign-problem-free bilayer extended Hubbard model}

\author{Runyu Ma}
\affiliation{School of Physics and Astronomy, Beijing Normal University, Beijing 100875, China\\}

\author{Zenghui Fan}
\affiliation{School of Physics and Astronomy, Beijing Normal University, Beijing 100875, China\\}

\author{Hongxin Liu}
\affiliation{School of Physics and Astronomy, Beijing Normal University, Beijing 100875, China\\}

\author{Tianxing Ma}
\email{txma@bnu.edu.cn}
\affiliation{School of Physics and Astronomy, Beijing Normal University, Beijing 100875, China\\}
\affiliation{Key Laboratory of Multiscale Spin Physics(Ministry of Education), Beijing Normal University, Beijing 100875, China\\}

\author{Hai-Qing Lin}
\affiliation{ School of Physics and Institute for Advance Study in Physics, Zhejiang University, Hangzhou 310058, Zhejiang, China \\}

\begin{abstract}
Competing orders represent a central challenge in understanding strongly correlated systems. In this work, we employ projector quantum Monte Carlo simulations to study a sign-problem-free bilayer extended Hubbard model. 
In this model, a charge stripe phase, characterized by a peak at momentum $k_x=2\pi\delta$ is induced by highly anisotropic interlayer spin-exchange coupling $J_z$, and strongly suppressed upon introducing the spin-flip term $J_\bot$; in contrast, \(J_\perp\) favors the emergence of interlayer pairing superconductivity. 
We further demonstrate that the anisotropy of the interlayer spin-exchange directly governs the competition between these two phases, while the on-site interaction \(U\) plays a complex role in tuning both the charge stripe and superconductivity. Our work identifies the key factors driving charge stripe formation, highlights the sensitivity of both the charge stripe and superconducting phases to interaction parameters, and thereby provides valuable insights into competing orders in strongly correlated systems.  
\end{abstract}

\maketitle

\section{Introduction}

One of the interesting phenomena in the Hubbard model is the competition between different phases\cite{RevModPhys.84.1383,PhysRevB.81.224505,RevModPhys.66.763}.
This also involves the crucial open problem of whether the Hubbard model can describe high-$T_{c}$ superconductors or not\cite{PhysRevB.37.5070,PhysRevB.37.7359,PhysRevLett.85.1524}.
Despite the simplicity of the pure Hubbard model, there are controversies even in the phase diagram since the solutions of the Hubbard model in more than one dimension in the strong coupling region
are hard to obtain\cite{doi:10.1146/annurev-conmatphys-031620-102024}.
Some experimental findings regarding high-$T_{c}$ superconductivity in cuprate-based materials may not be totally reproduced by the pure Hubbard model, especially the details of charge ordering and superconductivity\cite{doi:10.1126/science.1243479,Tranquada1995,Chang2012,doi:10.1146/annurev-conmatphys-031115-011401}.
One comprehensive study using the constrained-path quantum Monte Carlo and density matrix renormalization group algorithm reveals that in the pure Hubbard model under strong coupling and at a special electron filling, the stripe phase is dominant and superconductivity is suppressed\cite{PhysRevX.10.031016}. 
Furthermore, the periodicity of the charge stripe observed in experiments is not consistent with the numerical results from the pure Hubbard model\cite{doi:10.1126/science.aam7127}.
This may imply that some extra terms should be introduced into the pure Hubbard model to stabilize superconductivity and reproduce the correct charge ordering.

For example, long-range hopping can impair the stripe phase in some parameter regions and yied the correct periodicity of charge ordering\cite{doi:10.1126/science.aam7127, zhang2023frustrationinduced, doi:10.1126/science.aal5304, Huang2018}. Long-range hopping can also capture the doping asymmetry in experimental data.
The nearest neighbor interaction is another factor that is being investigated intensively.
Proposed by the investigations of one-dimensional cuprate chains\cite{doi:10.1126/science.abf5174},
the nearest-neighbor attractive interaction has been found to enhance the superconducting correlation and suppress the charge ordering\cite{PhysRevB.108.195136, PhysRevB.107.214504,PhysRevB.107.L201102,PhysRevB.105.155154}.
Other studies have introduced stripe order explicitly, which can be achieved by adding a spatially modulated potential.
The impacts of these modulations are complicated and it has been found that pairing strength and pairing symmetry can be affected\cite{chen2024charge,PhysRevB.109.045101,PhysRevB.98.121112}.
Furthermore, the charge-order correlation undergoes a $\pi$-phase shift when crossing a stripe\cite{PhysRevB.86.184506}.

Above all, the charge stripe phase is considered to be a very important feature that affects superconductivity in the Hubbard model.
A lot of work focuses on identifying factors that can suppress the stripe phase and thereby enhance superconductivity.
Understanding the charge stripe is of key importance in explaining unconventional superconductivity.
Various numerical methods have been developed to study Hubbard models, but most of them are only efficient under certain conditions\cite{PhysRevX.5.041041,PhysRevLett.109.186404,PhysRevLett.81.2514,PhysRevLett.69.2863,RevModPhys.78.865,RevModPhys.77.1027,PhysRevB.111.195127}.
For example, the determinant quantum Monte Carlo (DQMC) algorithm\cite{PhysRevD.24.2278} and its projector variant, the projector quantum Monte Carlo (PQMC), struggle with sign problem in many case of interest.

Sign problems occur in the quantum Monte Carlo algorithms when the Boltzmann factor is negative.
This can happen after decoupling the interaction terms with the Hubbard-Stratonovich (HS) transformation and tracing out the fermion degrees of freedom.
This is the main obstacle to overcoming the sign problem when using the DQMC and PQMC algorithms to study the rich physics in the Hubbard model.
One approach to overcome the sign problem is by utilizing sign-problem-free models. Among these models, the bilayer model, which belongs to the Kramers-class of sign-free models, shows potential for uncovering many-body effects in strongly correlated systems.
Previous work has found a transition from a Mott insulator to superconductivity, a transition analogous to that in cuprate superconductors\cite{PhysRevB.106.054510}.
Some other studies using a similar model have found a stable charge stripe phase with anisotropic interlayer interactions\cite{Assaad2002SpinAC,PhysRevB.108.165131}.
This demonstrates the value of this model in the investigation of stripe order phenomena, and also motivates us to undertake a more refined investigation of how interlayer interactions and on-site interactions influence stripe order.

In this work, we use the PQMC algorithm to compute the ground-state properties of a sign-problem-free bilayer Hubbard-like model with various parameters.
We find that the charge stripe phase is present when the interlayer spin-exchange is highly anisotropic.
Moreover, we find that this stripe phase is unstable upon introducing the spin-flip spin-exchange term $J_\bot$, and that an interlayer pairing superconductivity enhances when the stripe phase is suppressed.
Due to the flexibility of our model, we are able to tune the value of $U$ under certain range, $U\leq 1/4(J_{z} + 2J_{\bot})$), without encountering sign problems.
At small doping, the on-site interaction $U$ is detrimental to both the stripe phase and superconductivity. While the superconductivity is enhanced by $U$ at large doping, where charge stripe is absent.
We discuss the possible origin of these phenomena and their connections to the standard Hubbard model.

\section{Model and Method}
Our model was first proposed in Ref.~\cite{PhysRevLett.91.186402}, and subsequent works have found exotic phases and superconductivity within it\cite{PhysRevB.106.054510,PhysRevB.70.220505,PhysRevB.71.155115,PhysRevB.107.214509,2025-0936}.
This model originated from the $\rm{SO}(5)$ symmetric model\cite{PhysRevB.58.443}, and can be generalized to a bilayer model that contains both on-site interaction and interlayer spin-exchange. The Hamiltonian can be written as follows:
\begin{equation}
    \begin{aligned}
        H =& - t\sum_{\langle i, j\rangle l \sigma} \left( c^{\dagger}_{il\sigma} c_{jl\sigma} + H.c. \right) + U \sum_{i} n_{i1\uparrow} n_{i2\downarrow}\\
        &+J_{z} \sum_{i} S^{z}_{i1} S^{z}_{i2} + \frac{J_{\bot}}{2} \sum_{i} \left( S^{+}_{i1} S^{-}_{i2} + H.c. \right)
    \end{aligned}
    \label{eq:ham}
\end{equation}
where $c_{il\sigma}$($c^{\dagger}_{il\sigma}$) denotes the annihilation (creation) operator at site $i$, layer $l$, and spin $\sigma$.
$n_{il\sigma} = c^{\dagger}_{il\sigma}c_{il\sigma}$ is density operator, $S^{z}_{il} = \frac{\left(n_{il\uparrow} - n_{il\downarrow}\right)}{2}$ is $z$ component spin operator,
and $S^{+}_{il} = c^{\dagger}_{il\uparrow}c_{il\downarrow}$ (its Hermitian conjugate $S^{-}_{il}$) is spin annihilation (creation) operator, contributing to the $xy$ component spin.

The absence of the sign problem in this model originates from time-reversal symmetry. To illustrate this, we define 
$\psi_{i} = \left(c_{i1\uparrow}, c_{i1\downarrow}, c_{i2\uparrow}, c_{i2\downarrow}\right)^{T}$, and
rewrite the Hamiltonian in the following form:
\begin{equation}
    \begin{aligned}
        H = & -t \sum_{\langle i, j \rangle} \left( \psi^{\dagger}_{i} \psi_{j} + H.c. \right) - \sum^{5}_{\alpha=1} \frac{g_{\alpha}}{2} \left(\psi^{\dagger}_{i} \Gamma^{\alpha} \psi_{i}\right)^2  \\
        & - \frac{g_{0}}{2} \sum_{i} \left(\psi^{\dagger}_{i} \psi_{j} -  2\right)^2
    \end{aligned}
    \label{eq:ham2}
\end{equation}
where we define
\begin{equation}
    \Gamma^{1\rm{-}3} = \begin{bmatrix}
        \mathbf{\sigma} & 0 \\
        0 & -\mathbf{\sigma}
    \end{bmatrix} \quad
    \Gamma^{4} = \begin{bmatrix}
        0 & I \\
        I & 0
    \end{bmatrix} \quad
    \Gamma^{5} = \begin{bmatrix}
        0 & iI \\
        -iI & 0 
    \end{bmatrix}
\end{equation}
and $\mathbf{\sigma}$ represents the Pauli matrices. In this work, we focus on the case where $g_{1}=g_{2}$ and $g_{4}=g_{5}$, and
\begin{equation}
    \begin{aligned}
        g_{0} &= -\frac{3}{4} V - \frac{1}{4} U + \frac{1}{8} J_{\bot} + \frac{1}{16} J_{z} \\ 
        g_{1,2} &= -\frac{1}{4} V + \frac{1}{4} U + \frac{1}{8} J_{\bot} - \frac{1}{16} J_{z} \\
        g_{3} &= -\frac{1}{4} V + \frac{1}{4} U - \frac{1}{8} J_{\bot} + \frac{3}{16} J_{z} \\
        g_{4,5} &= \phantom{-} \frac{1}{4} V - \frac{1}{4} U + \frac{1}{8} J_{\bot} + \frac{1}{16} J_{z} \\
    \end{aligned}
\end{equation}

After we map the original model Eq.(\ref{eq:ham}) to Eq.(\ref{eq:ham2}), we focus on the five $\Gamma^{\alpha}$ matrices defined above.
It has been proved that these matrices are invariant under the time-reversal transformation $\mathcal{T}  = R\mathbb{C}$, where $\mathbb{C}$ denotes complex conjugation and $R$ is given by
\begin{equation}
    R = \begin{bmatrix}
        0 & -i \sigma_2\\
        -i\sigma_2 & 0
    \end{bmatrix}
\end{equation}
This ensures that the fermion determinant is positive. For details, see Ref.\cite{PhysRevLett.91.186402} and Ref.\cite{PhysRevB.71.155115}.

In this work, we use the PQMC method, through which the observables are calculated by the projection as follows:
\begin{equation}
    \begin{aligned}
        \langle O \rangle = \frac{\langle \Phi_{T} \vert e^{-\Theta H} O e^{-\Theta H} \vert \Phi_T \rangle}{\langle \Phi_{T} \vert e^{-2\Theta H} \vert \Phi_{T} \rangle}
    \end{aligned}
    \label{eq:pro}
\end{equation}
To treat the quantum system, we first utilize the Trotter decomposition, $e^{-\Theta H} = \left( e^{-\Delta \tau H} \right)^{N_{t}} = \prod^{N_{t}}_{\tau} e^{-\Delta \tau V} e^{-\Delta \tau T}$,
where $N_{t}\Delta\tau = \Theta$, $T$ is the hopping term $-t\sum_{\langle i, j \rangle} \left( \psi^{\dagger}_{i} \psi_{j} + H.c. \right)$, and $V$ represents the remaining terms in Hamiltonian Eq.(\ref{eq:ham2}). Each term in $V$
is a perfect square of bilinear operator. We decouple these terms using the standard HS transformation:
\begin{equation}
    \begin{aligned} 
e^{-\Delta \tau V} &= e^{\sum^{5}_{\alpha=0} \sum_{i} \frac{\Delta \tau g_{\alpha}}{2} \left( \psi^{\dagger}_{i} \Gamma^{\alpha} \psi_{i} - z_{\alpha} \right)^2 } \\
&= \prod_{\alpha,i} \sum_{k=\pm 1, \pm 2} \gamma_{k,i} e^{\sqrt{\Delta \tau g_{\alpha}}  \eta_{k,i} \left( \psi^{\dagger}_{i} \Gamma^{\alpha} \psi_{i} - z_{\alpha} \right)} \\
&= \sum_{k} \prod_{i} \gamma_{k,i} U(k,i)
    \end{aligned}
    \label{eq:HS}
\end{equation}
where $z_{0} = 2$ and $z_{1\rm{-}5} = 0$. The coefficients $\gamma$ and $\eta$ follow the standard form for the four-component HS transformation, with details available in Ref.\cite{goth2020higher}.
After the HS decomposition, the system transforms into an ensemble of free fermions. The Green's functions for a given configuration can be calculated easily, thereby rendering the observables a weighted average over these free-fermion configurations. The weight of each configuration is defined as $\omega  = \prod \gamma^{\tau}_{k} \det \left[P^{\dagger} B_{V}(\vec{k}_{N_{t}})B_{T} \ldots B_{V}(\vec{k}_{1}) B_{T} P\right]$, where $B_{T}$ denotes the matrix form of $e^{-\Delta \tau T}$ in the single-particle basis, 
and $B_{V}(\mathbf{s}_{\tau})$ is the matrix form of $\prod_{i} U(k,i)$ for a specific configuration $\vec{s}_{\tau}$ at imaginary time $\tau$ in the single-particle basis. $P$ is a matrix that characterizes the Slater determinant of the trial wave function $\Phi_{T}$.
The weight $\omega$ is crucial for PQMC simulations, and the sign problem emerges when its value is non-positive. 
Fortunately, if the matrix in the determinant remains invariant under the $\mathcal{T}$ transformation, its determinant is consistently positive. 
The model employed in this work satisfies this condition, allowing us to perform quantum Monte Carlo simulations free of the sign problem.

\section{Results and Discussions}
When the interlayer interactions are purely anisotropic, previous work has shown that this model exhibits both
charge and spin stripe phases\cite{Assaad2002SpinAC, PhysRevB.108.165131}. Moreover, this state remains robust even when next-nearest-neighbor hopping is incorporated.
To characterize the charge stripe, we define the charge correlation function as follows:
\begin{equation}
    \begin{aligned}
        S_{ch}\left(\mathbf{k}\right) = \frac{1}{N} \sum_{i,j} e^{-i\mathbf{k}\cdot \left(\mathbf{R}_{i} - \mathbf{R}_{j}\right)}
        \left[ \langle n_{i} n_{j} \rangle - \langle n_{i} \rangle \langle n_{j} \rangle \right]
    \end{aligned}
\end{equation}
In the following, we present the charge correlation function at $k_{y}=0$, where the peak position along $k_{x}$ reflects the periodicity of charge modulations. 
Unless otherwise specified, we use a system size with $L_y = 8$ and $L_x = 2/\delta$, where $\delta$ denotes the hole doping density.

\begin{figure}
    \centering
    \includegraphics[width=0.48\textwidth]{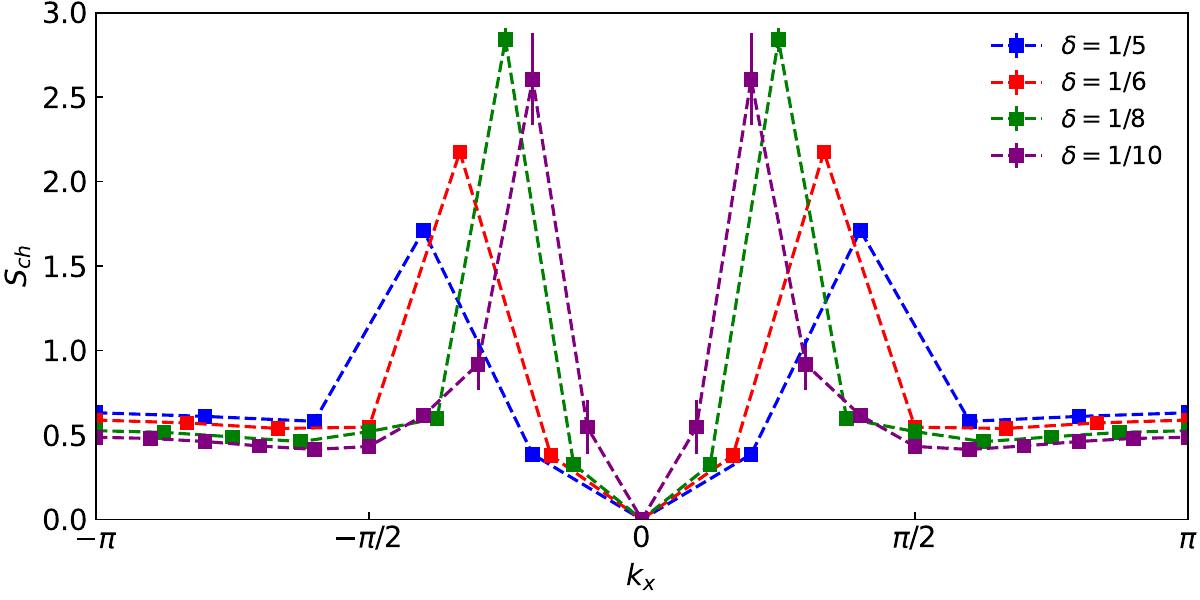}
    \captionsetup{justification=raggedright}
    \caption{The charge correlation function $S_{ch}(\mathbf{k})$ from momentum $k=(-\pi, 0)$ to $k=(\pi, 0)$ for different hole doping densities $\delta$ at on-site interaction $U=2.0$ with pure z-component interlayer spin-exchange $J_{z}=4U$.
    The peak indicates charge stripe order. We fix $L_y=8$ and $L_x = 2/\delta$ in our simulations.}
    \label{fig:cdwlx}
\end{figure}

As shown in Fig.\ref{fig:cdwlx}, when the system contains only the $U$ and $J_{z}$ terms, the peak in $S_{ch}(k_{x}, 0)$ is at $2\pi \delta$. The peak intensity of $S_{ch}$ is maximal at $1/8$ doping, a feature consistent with previous work on cuprate superconductors\cite{Huang2018, PhysRevResearch.2.033073}.
This phenomenon is interesting because the sign-problem-free model we employ  shows stripe behavior similar to that in cuprate superconductors, despite the different geometries of the two systems. This suggests that the underlying mechanisms might be similar.
Previous work (Ref.\cite{PhysRevB.108.165131}) also reveals that this model shares another remarkable similarity with the single-layer Hubbard model, and the introduction of a next-nearest-neighbor hopping term shortens the stripe wavelength.

\begin{figure}
    \centering
    \includegraphics[width=0.48\textwidth]{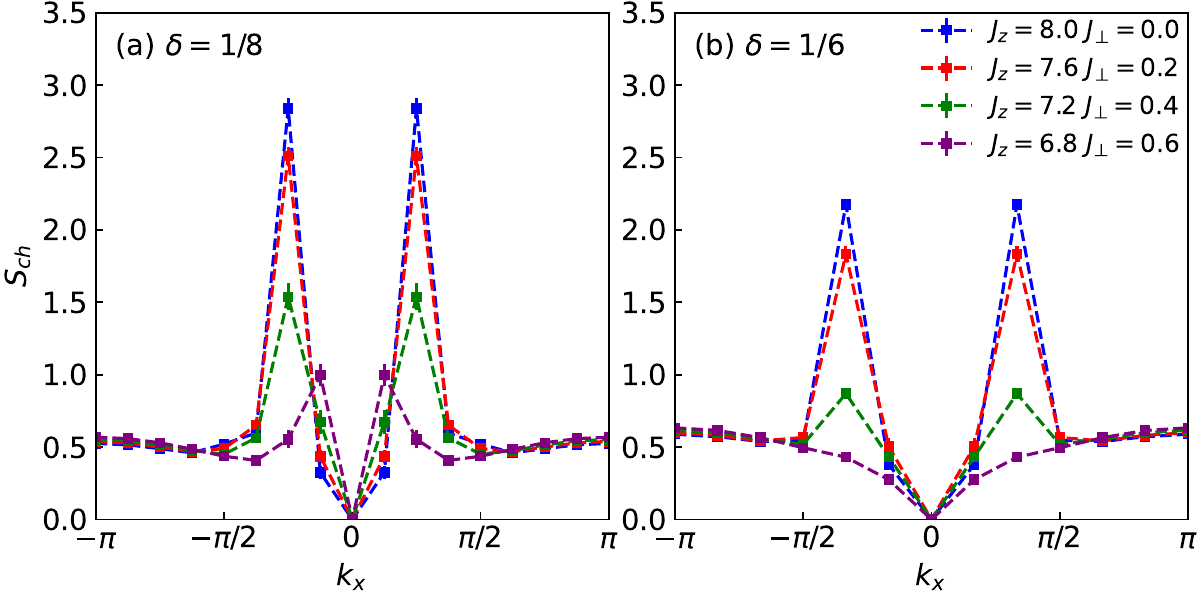}
    \caption{The charge correlation function $S_{ch}(k_{x},0)$ at $U=2.0$ for different $J_{\bot}$ and $J_{z}$ in the system of (a) $L_{x}=16$, $L_{y}=8$ at $1/8$ doping and (b) $L_{x}=12$, $L_{y}=8$ at $1/6$ doping.}
    \label{fig:jbot}
\end{figure}

We now take a deeper look at the stripe phase.
Since the interlayer Ising term $J_{z}$ favors antiparallel spin pairs on a rung, holes introduced into the system do not spread homogeneously because doing so would break these spin pairs.
Holes instead concentrate on specific sites to minimize the system’s energy.
However, this order is readily disrupted by introducing fluctuations to the interlayer spin-exchange; namely, adding the $J_{\bot}$ term suppresses  the stripe phase.
To maintain this model as sign-problem-free, the relation $J_{z} + 2J_{\bot} = 4U$ must be satisfied.
With $U=2.0$ fixed, we tune the ratio of $J_{z}$ to $J_{\bot}$ and observe that this peak disappears rapidly, as shown in Fig.~\ref{fig:jbot}. 

\begin{figure}[b]
    \centering
    \includegraphics[width=0.48\textwidth]{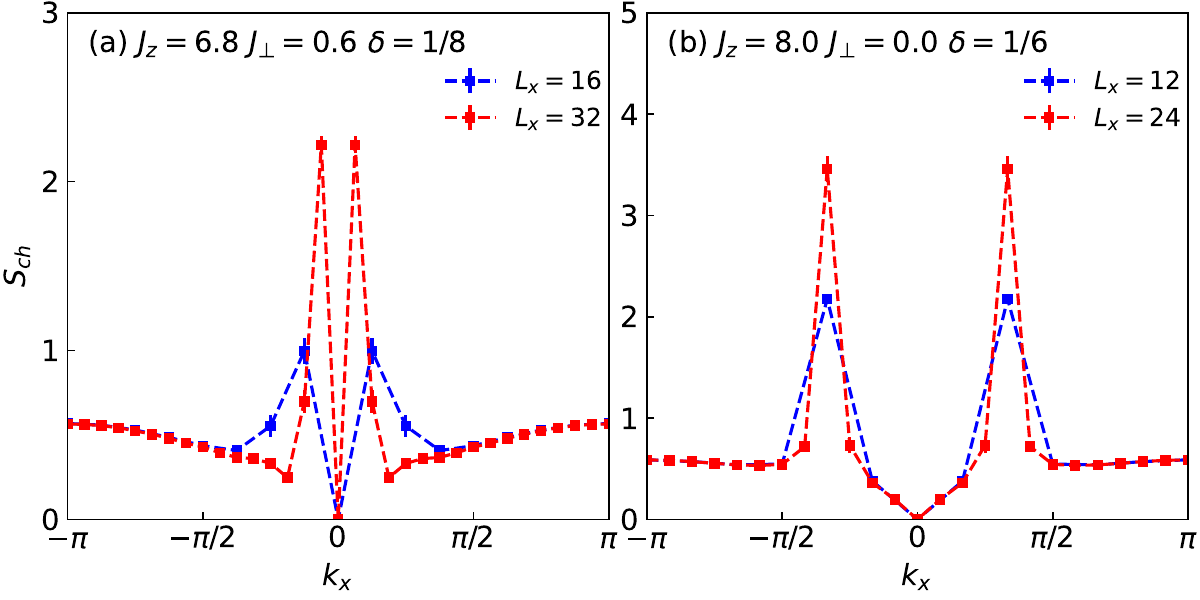}
    \caption{The charge correlation function $S_{ch}(k_{x},0)$ for $U$=2 with relationship $J_{z}+2J_{\bot}=4U$: (a) $J_{z}$ with $J_{\bot}$ case, $\delta = 1/8$ with $L_{x}=32$ and $L_{x} = 16$; (b) pure $J_{z}$ case, $\delta = 1/6$ with $L_{x}=24$ and $L_{x} = 12$.}
    \label{fig:largex}
\end{figure}

When we introduce $J_{\bot}$ at $\delta=1/8$, we observe that the peak position shifts, as shown in Fig.~\ref{fig:jbot}.
The peak moves to $\pi / 8$, which corresponds to a charge period that matches the system size $L_x$ given the condition $L_x = 2/\delta$. 
To allow $L_x$ to accommodate more than one period of charge modulation, we increase $L_x$ to 32, and the peak position shifts to a smaller wavevector, as shown in .
Figure~\ref{fig:largex}.
The discrepancy between results for different system sizes points to the influence of finite-size effects for large $J_{\bot}$.
Due to the finite-size effects, the periodicity of the stripe is spurious, and the stripe may be absent in the thermodynamic limit. 
Details are provided in the discussion about Fig.~\ref{fig:largex} below.
However, due to the constraints on computational power, further verification on this issue is exceedingly difficult.

\begin{figure}[b]
    \centering
    \includegraphics[width=0.48\textwidth]{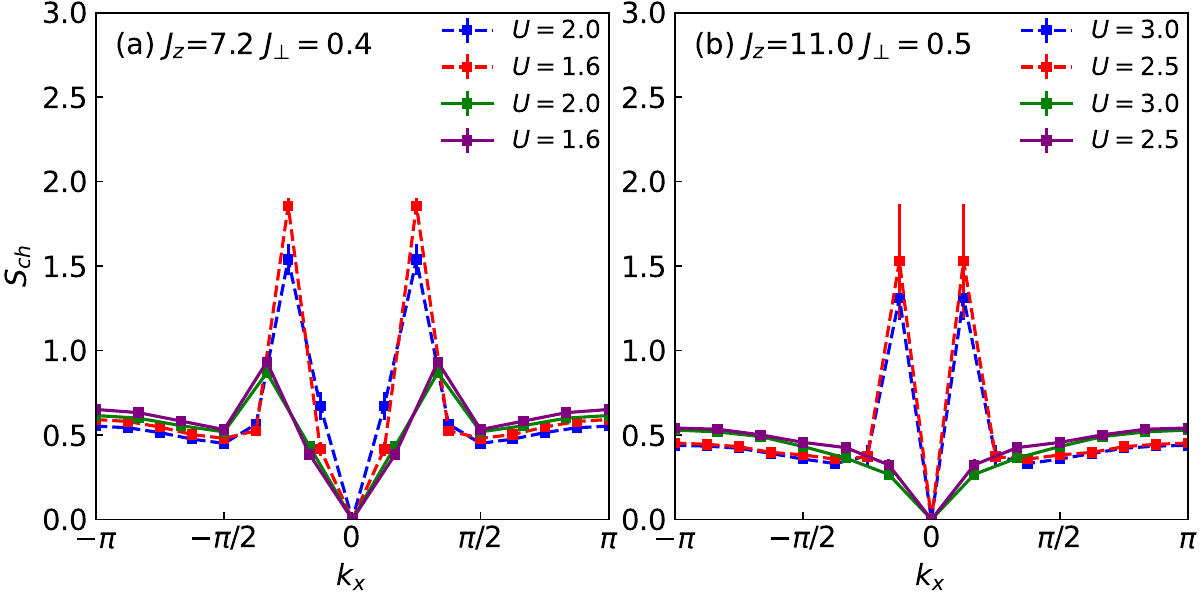}
    \caption{The charge correlation function $S_{ch}(k_{x},0)$ for different values of $U$ with doping $\delta = 1/8$ (dashed line) and $\delta = 1/6$ (solid line) at (a) $J_{z} = 7.2$ and $J_{\bot}=0.4$ and (b) $J_{z} = 11.0$ and $J_{\bot}=0.5$.
    }
    \label{fig:difU}
\end{figure}

In the single-layer Hubbard model, a comparable spin-exchange interaction arisess from a perturbative treatment of electron hopping,
leading to an effective $t$-$J$ model\cite{PhysRevB.18.3453, tian2019mapping}. 
The stripe state in the single-layer model maintains the energetically favored antiferromagnetic order between stripes.
In this sense, the mechanism underlying stripe formation in the two models is similar.
Spin-exchange in the single-layer lattice also involves attraction between antiparallel spins. 
The presence of spin-exchange between four nearest-neighboring sites can induce a larger energy difference. Consequently, the stripe phase in the single-layer lattice is expected to be more stable than that in the bilayer lattice, where only interlayer spin-exchange exists.
Furthermore, Ref.~\cite{PhysRevB.106.195121} studied nonlocal interactions in the extended Hubbard model and found that nearest-neighbor attraction enhances antiferromagnetic correlations.
To conclude this section, the stripe phase in the two-dimensional Hubbard model is likely more stable, which reflects the difference between our sign-free model and the pure Hubbard model.

\begin{figure}
    \centering
    \includegraphics[width=0.48\textwidth]{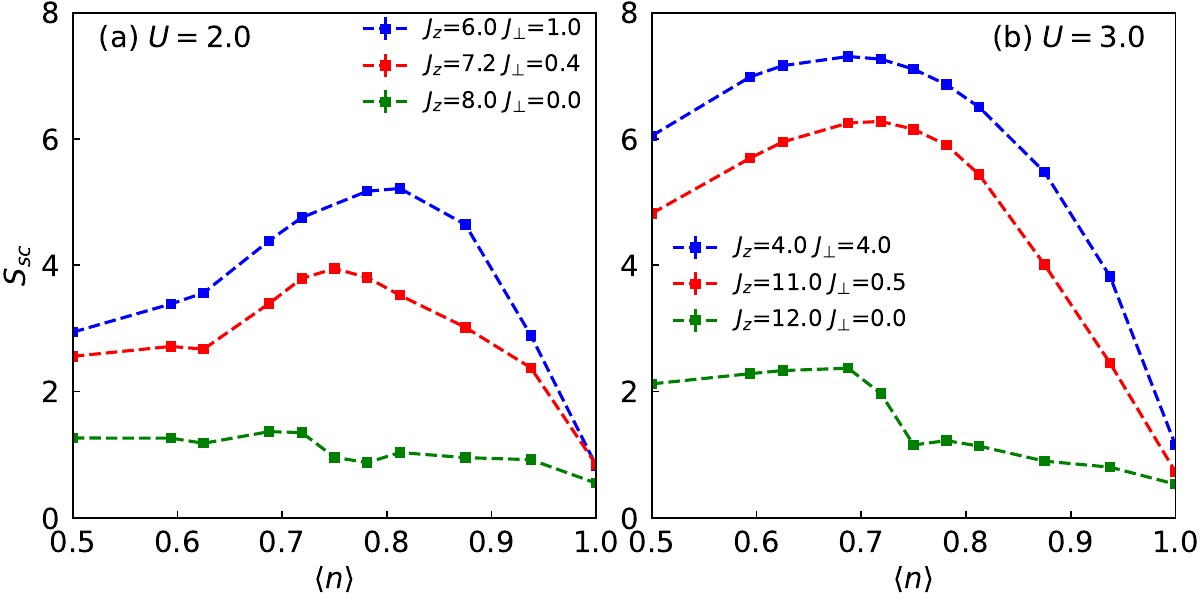}
    \caption{The pairing correlation function $S_{sc}$ as a function of electron filling $\langle n\rangle$ for different values of $J_{z}$ and $J_{\bot}$ at (a) $U=2.0$ and (b) $U=3.0$. The system size is $L_{x}=8$ and $L_{y}=8$.}
    \label{fig:esvn}
\end{figure}

Next, we gradually suppress the $U$ term in the Hamiltonian under the condition of $J_{\bot} \ne 0$. 
To avoid the sign problem, the value of $U$ must be larger than $\frac{J_{z}}{4} - \frac{J_{\bot}}{2}$. 
Given that the on-site interaction $U$ prohibits double occupation, it tends to favor the formation of 
antiparallel spins. 
However, $U$ also favors other ordered phases, including superconductivity and in-plane magnetism. 
The involvement of possible intralayer order constitutes another difference between this model and the single-layer Hubbard model.
As shown in Fig.~\ref{fig:difU}, increasing $U$ suppresses the stripe phase, which suggests that $U$ may be more inclined towards intralayer order. 
Moreover, as observed in Fig.~\ref{fig:difU} (b), the charge stripe vanishes when the doping increases to $\delta = 1/6$ for $J_z=11.0$ and $J_{\bot}=0.5$. 
Compared with Fig.~\ref{fig:difU} (a), this indicates a narrower doping range around the optimal $\delta = 1/8$ at larger $U$ --- a reflection of $U$-induced suppression of the charge stripe, where the suppressive effect of $U$ outweighs the facilitative effect of $J_z$. 

\begin{figure}
    \centering
    \includegraphics[width=0.48\textwidth]{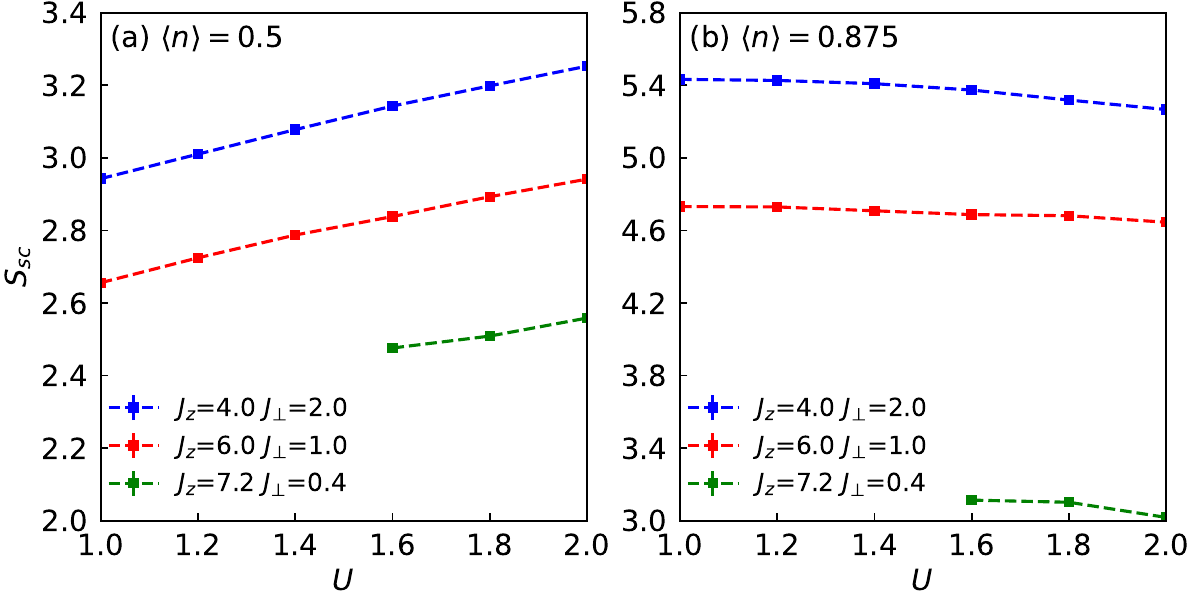}
    \caption{The pairing correlation function $S_{sc}$ as a function of electron interaction $U$ for fixed values of $J_{z}$ and $J_{\bot}$ at (a) $\langle n \rangle = 0.5$ and (b) $\langle n \rangle = 0.875$. The system size is $L_{x}=L_{y}=8$.
    }
    \label{fig:esvU}
\end{figure}

The superconductivity in this system also shows a complex dependence on doping and interaction strength.
We focus on the interlayer pairing and its correlation functions which can be defined as
\begin{equation}
    S_{sc} = \frac{1}{N} \sum_{i,j} \langle \Delta^{\dagger}_{i} \Delta_{j}  \rangle
\end{equation}
where $\Delta_{i} = c_{i1\uparrow} c_{i2\downarrow} - c_{i1\downarrow} c_{i2\uparrow}$.

Now we present the pairing correlation function as a function of electron filling $\langle n \rangle$ in Fig.~\ref{fig:esvn}.
This figure clearly reveals the evidence of competition between charge stripe and superconductivity. 
On the one hand, the isotropy of $J$ (i.e., introducing $J_{\bot}$) enhances superconductivity by disrupting the charge stripe. 
On the other hand, for the pure $J_{z}$ case,  superconductivity decreases sharply around 1/4 doping where charge stripe emerges
for both $U$=2 and $U$=3. This abrupt change is particularly pronounced for $U=3$. 
Thus, for the pure $J_{z}$ case, the pairing correlation function exhibits a flat profile below 1/4 doping, attributed to suppression by charge stripes. In contrast, in the presence of $J_{\bot}$, the pairing correlation function shows a smooth increase as the doping level rises to 1/4, since charge stripes are weakened and gradually disappear in this region. This provides clear evidence for the competition between charge stripes and superconductivity.

However, the relationship between superconductivity and charge stripes in this model is complex.
We gradually decrease $U$ when $J_{\bot}$ is present, as shown in Fig.~\ref{fig:esvU}. Due to the constraints on the model parameters, a smaller $U$ can be achievable when $J{\bot}$ is large.
We previously found that the charge stripes are suppressed as $U$ increases, and we now observe that superconductivity is also suppressed at low doping levels with increasing $U$, as shown in Fig.~\ref{fig:esvU}(b).
This provides another piece of evidence for the influence of intralayer ordered phases.
At lower electron densities, where charge stripes do not form, increasing $U$ enhances superconductivity.
This reflects the multiple roles played by parameters $U$ and $J$,
while the electron density also plays a significant role in regulating the system's properties..

Finally, the properties of the results for larger system sizes are verified by increasing $L_x$. 
Figure~\ref{fig:largex} (b) presents the case with pure $J_{z}$, which shows that the peak position remains unchanged while its intensity increases significantly as $L_x$ increases. 
This intensity enhancement may suggest the emergence of long-range order, which requires further investigation ---a pursuit that is limited in this work due to computational constraints. In contrast, as previously mentioned for Fig.~\ref{fig:jbot} (a), the peak for $J_z=6.8$ shifts toward a smaller wavevector $\pi/8$ in momentum space compared to the pure $J_z$ case, indicating only one period in the $L_x=16$ system. After increasing $L_x$ to 32 as shown in Fig.~\ref{fig:largex} (a), the peak position shifts further to a smaller wavevector compared with that in the $L_x=16$ case, while the system still contains only one stripe period. This may imply that as $L_x\rightarrow \infty$, the peak will shift to zero, which corresponds to an infinite stripe periodicity (i.e., a uniform density distribution). This suggests that the observed charge stripe for this parameter set may be a finite-size effect, with no stripe order existing in the thermodynamic limit.

\section{Conclusions}

Using the PQMC algorithm, we study a sign-problem-free bilayer Hubbard-like model to investigate the competitive interplay between charge stripes and interlayer superconductivity, which is controlled by interlayer spin-exchange anisotropy ($J_z$ and $J_{\bot}$) and on-site interaction $U$. A charge stripe phase, analogous to that in the pure two-dimensional Hubbard model, emerges in the bilayer model. We find that interlayer spin-exchange plays a crucial role in the formation of both charge stripes and interlayer superconductivity. The charge stripe phase, which is characterized by a peak at $k_x=2\pi\delta$, is stable for highly anisotropic coupling $J_z$ but strongly suppressed with the introduction of the spin-flip term $J_\bot$. Concurrently, superconducting correlations are enhanced as charge stripes diminish, indicating a direct competition between these two phases. We also discuss the possible relationship between this model and the pure two-dimensional Hubbard model, in which such competition also exists. Furthermore, the on-site interaction $U$ plays a complex role: As $U$ increases, it suppresses both charge stripes and superconductivity under low doping, suggesting the emergence of other intralayer orders. However, with further increases in doping, increasing $U$ enhances superconductivity in the absence of charge stripes. This work identifies the key factors contributing to the formation of charge stripes, and highlights the sensitivity of charge stripe and superconducting phases to interaction parameters, thus offering insights into competing orders in strongly correlated systems.

\section{Acknowledgements}

This work is supported by NSFC (12474218) and Beijing Natural Science Foundation (No. 1242022 and 1252022). 
The numerical simulations in this work were performed at the HSCC of Beijing Normal University.

\bibliography{reference}

\end{document}